# Polarization dependent chemistry of ferroelectric BaTiO$_3$ (001) domains

*Yanyu Mi[1], Gregory Geneste[2,3]\*, Julien Rault[1], Claire Mathieu[1], Alexandre Pancotti[1], Nicholas Barrett[1]*

[1]IRAMIS/SPCSI/LENSIS, F- 91191 Gif sur Yvette cedex, France.

[2]CEA, DAM, DIF, F-91297 Arpajon, France.

[3]Laboratoire Structures, Propriétés et Modélisation des Solides, CNRS-UMR 8580, Ecole Centrale Paris, Grande Voie des Vignes, 92295 Chatenay-Malabry Cedex, France

\*gregory.geneste@cea.fr

ABSTRACT. Recent works suggest that the surface chemistry, in particular, the presence of oxygen vacancies can affect the polarization in a ferroelectric material. This should, in turn, influence the domain ordering driven by the need to screen the depolarizing field. Here we show using density functional theory that the presence of oxygen vacancies at the surface of BaTiO$_3$ (001) preferentially stabilizes an inward pointing, P$^-$, polarization. Mirror electron microscopy measurements of the domain ordering confirm the theoretical results.

I - Introduction

A fundamental property of ferroelectric (FE) materials is their electrically switchable spontaneous polarization below the Curie temperature, which has driven promising applications of such materials as nonvolatile memory storage devices [1]. However, the direction of the polarization in FE thin films is not only the result of a simple control through external electric field, since it usually results from the minimization of the electrostatic energy (driven by an equilibrium between short and long-range forces) in the whole sample [1]. Moreover, it has been recently shown [3] that the chemical environment has strong interaction with the polarization orientation of perovskite ferroelectric oxides: on the one hand, the polarization orientation can influence surface chemical reactions (the adsorption of various molecules depends on the polarization direction), and on the other hand, the chemical composition of the atmosphere (the oxygen chemical potential for instance) might induce preferential orientation of the polarization [4]. It has been suggested that oxygen vacancies play a role in the stabilization of negative polarization, i.e. polarization pointing inwards [5]. Recent work has also investigated the coupling between ferroelectricity and adsorbed water on $BaTiO_3$(001) surfaces [13,14].

Oxygen vacancies are common point defects in perovskite ferroelectric oxides. They are considered to be directly related with the instability of ferroelectric polarization, resulting in ferroelectric fatigue, and might be associated to various charge compensation mechanisms. However, the exact role played by such defects in the ferroelectric switching process, and the effect of their surface chemistry on ferroelectric domains with different polarization orientations is still an open question. A study integrating the chemical state of ferroelectric oxide surfaces with different polarization orientation is expected to add to a fundamental understanding of the ferroelectric behavior and related degradation mechanisms.

This letter is devoted to $BaTiO_3$ (001) (BTO(001)) single crystal surfaces with different spontaneous polarization orientations. We apply both a theoretical approach based on density-functional calculations and an experimental approach using mirror electron microscopy with a low energy electron microscope. We provide insight into the link between polarization orientation and surface chemistry.

In the following, c$^+$ (resp. c$^-$) are domains with polarization pointing outwards (resp. inwards) and denoted P$^+$ (resp. P$^-$), while "a" domains have their polarization lying in-plane.

**II - Density-Functional calculations**

Simulating ferroelectric surfaces with uniform polarization perpendicular to the surface raises technical problems related to the fact that in such configuration, a strong depolarizing electric field generated by the surface charges usually suppresses the ferroelectricity. Despite this fact, ferroelectric thin films often exhibit such configurations. However, in such cases, alternating domains with "up" and "down" dipole moments are usually observed [17]. Rather than making the FE disappear (or lie parallel to the surface), the formation of such alternating domains is the way the system finds to minimize its electrostatic energy. If a domain with "up" polarization (for example) can exist, it means that the corresponding depolarizing field (acting inside the domain) is exactly compensated by another "external" macroscopic electric field. This external field is the one provided (mainly) by adjacent domains, which both have "down" polarization. Let us point out that this is an "external" field from the point of view of the domain, although it is produced by the system itself and not by an external operator. A ferroelectric domain with polarization perpendicular to the surface can thus, in principle, be simulated by reproducing this external field [18], just as in reality.

We perform density-functional calculations using the SIESTA code [15,16]. We use the Local Density Approximation (LDA) and Troullier-Martins pseudopotentials. A basis of numerical atomic orbitals is used, extended up to triple dzeta in the case of Ti 3d and 4s, and O 2p. Simulation of the oxygen vacancy is achieved by placing a ghost atom at the defect site. The fineness of the real space grid is determined by a cut-off of 400 Rydberg and the range of atomic orbitals by an energy shift of 0.001 Rydberg.

We construct 9-layer (001)-oriented slabs with BaO termination on both sides, and apply an external electric field **E$_{ext}$** along the [001] direction, i.e. perpendicular to the surface. This external field is associated to a sawtooth electrostatic potential having its discontinuity in the middle of the vacuum. A large vacuum of more than 20 Å separates the periodically repeated slabs. First, perfect surfaces have been calculated, using a 1x1 surface unit cell (the Brillouin zone is sampled by a 6x6x1 mesh). Then the surface unit cell is doubled in the two in-plane directions (the Brillouin zone of the supercell is accordingly sampled by a 3x3x1 mesh), and an oxygen vacancy is placed on the bottom surface plane (the final system contains 87 atoms). If $E_{ext} > 0$ (resp. $< 0$), the system will be denoted by "P$^-$" (resp. "P$^+$") and the bottom surface of the slab has a negative (positive) fixed polarization charge. The geometry of the system is optimized using a conjugate-gradient algorithm until the maximal Cartesian component of atomic forces is smaller than 0.04 eV/Å. Two lateral lattice constants have been tested: the LDA lattice constant of cubic BTO (3.940 Å) and another one closer to the experimental value (3.995 Å).

The calculations performed on the perfect surface allow us to find the value of the external field that produces a polarization roughly equal to the spontaneous polarization of BTO: we find ~ 1.2 V/Å. Thus, we perform four series of calculations (for each lateral lattice constant and each external field direction P$^+$ and P$^-$), by gradually increasing the applied external field from zero up to 1.4 V/Å by step of 0.2 V/Å. Fig. 1 shows one of the "P$^+$" systems after geometric optimization.

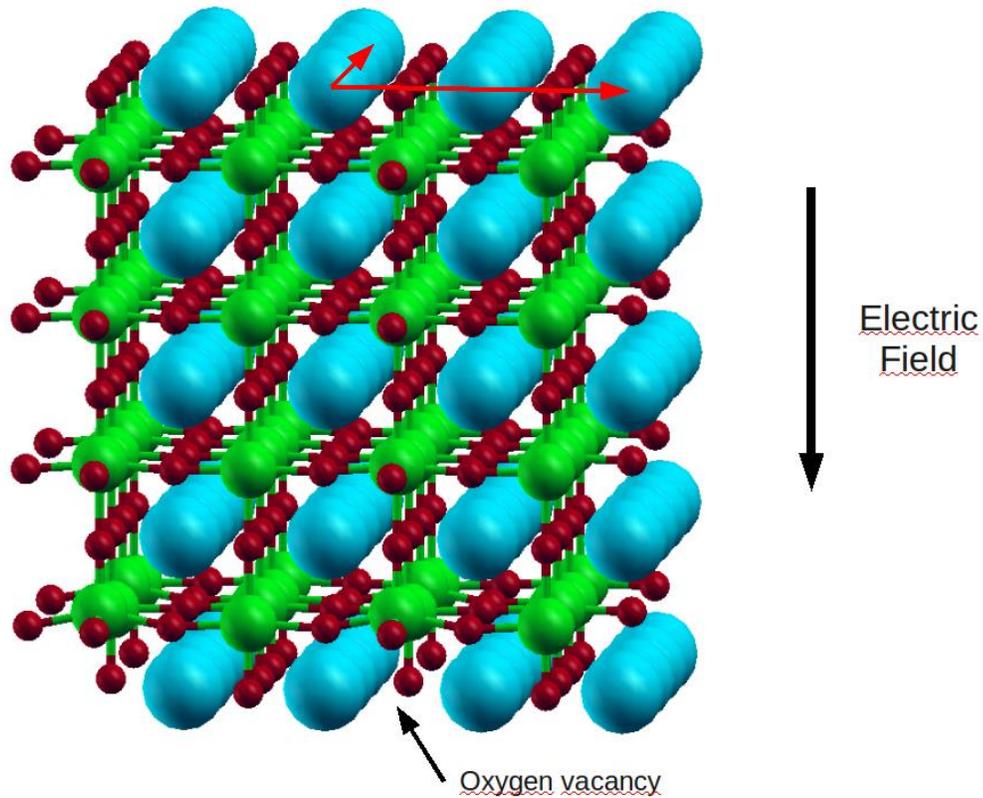

"**Figure 1.** Figure caption. Side view of the 9-layer slab used in the present calculations, in the case of an external electric field pointing downwards, the oxygen vacancy being at the bottom surface ($P^+$). Large blue, medium green and small red balls are respectively Ba, Ti and O atoms. The red arrows highlight the 2x2 surface unit cell."

Fig. 2 presents the difference of energy between $P^-$ and $P^+$ as a function of applied external electric field, for the two lateral lattice constants studied. In both cases, and for any external field, the quantity $E(P^-) - E(P^+)$ is found negative, showing that $P^-$ is more stable than $P^+$ under similar conditions of external electric field. Note that this energy includes the coupling between the polarization and the field.

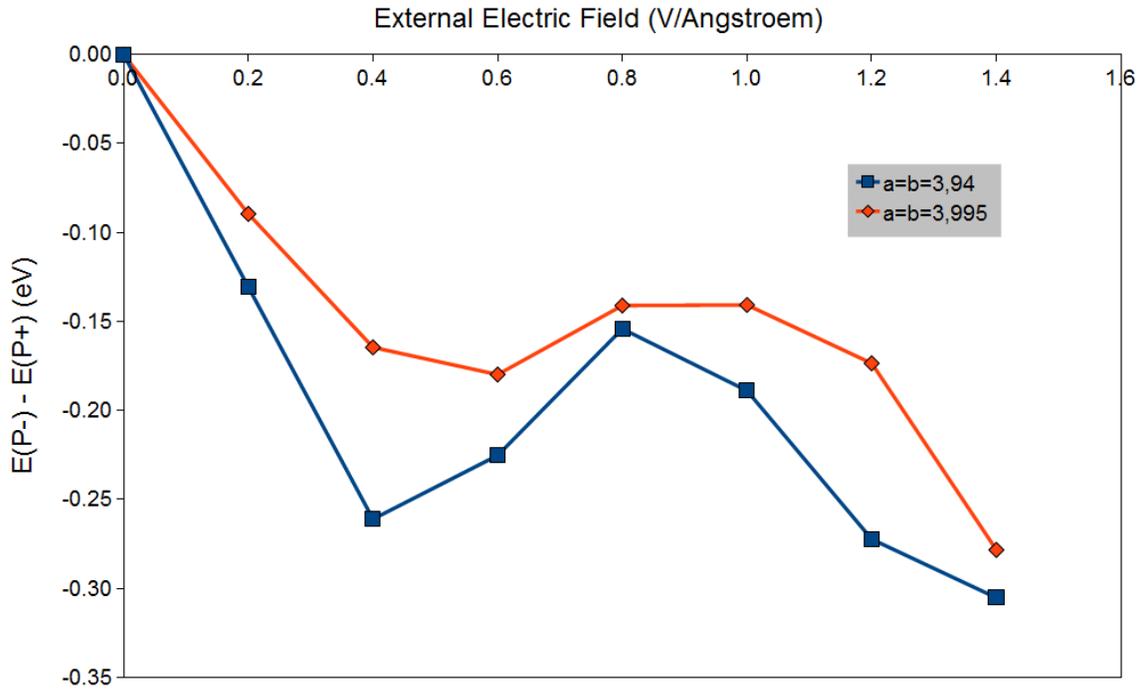

"**Figure 2.** Figure caption. $E(P^-) - E(P^+)$ (eV) as a function of applied external electric field (V/Å)."

Analysis of electronic populations and atomic magnetic moments shows that the Ti atoms of the $TiO_2$ plane closest to the bottom surface containing the vacancy acquire a small magnetic moment whereas all the others remain in a non-magnetic state. The Mulliken population of the Ti next to the oxygen vacancy slightly increases, consistently with the donor character of the neutral oxygen vacancy. Of course, such calculations are only relevant to compare the stability of free-standing slabs, i.e. with one surface exhibiting a vacancy and another one without defect, under similar external field.

**III - Experiments**

So far, most studies on the ferroelectric domains are done by scanning probe based techniques, such as piezoresponse force microscopy (PFM) [6,7] and electrostatic force microscopy (EFM) [8]. Mirror electron microscopy (MEM), i.e. operating a Low energy electron microscope (LEEM) in "mirror" mode, is sensitive to the electrostatic potential just above the sample surface [9] which is in turn related to the fixed polarization charge at the domain surface. Although the electrostatic potential above the surface is ill-defined, (for example in a perfect ferroelectric slab, the magnitude of the potential depends on the slab thickness), MEM can be used for weak interaction imaging providing that the domain width is much greater than the domain wall width, as is usually the case. Furthermore, it avoids artifacts introduced by tip-surface interactions inherent to scanning probe microscopy [10]. Of course, in the case of complete screening of the polarization charge there should be no observable difference in the electrostatic potential above $c^+$, $c^-$ and in-plane domains but if the system energy is minimized through domain ordering then MEM should be a sensitive probe of the latter.

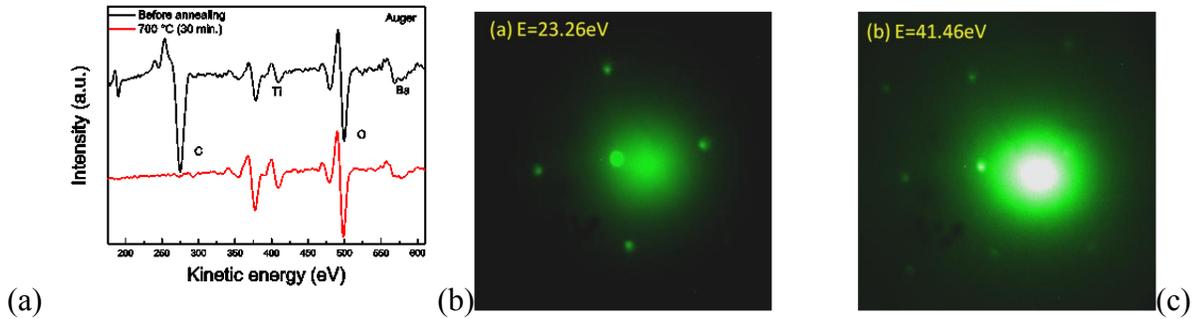

"**Figure 3.** Figure caption. (a) AES of BTO(001) surface before and after in-situ annealing at 700°C LEED pattern of BaTiO$_3$(001) sample taken with the LEEM with primary energy of 23.26 eV (b) and 41.46eV (c), respectively."

The as-received BaTiO$_3$ (001) single crystal (SurfaceNet GmbH) is annealed in O$_2$ at 4.5 mbar for 1 hour at 873°C to obtain a clean surface reconstruction. To eliminate surface carbon contamination, the sample is cleaned by UV-ozone just before being introduced into the UHV chamber of an Elmitec III LEEM. In-situ annealing to ensure an adsorbate free surface and to create oxygen vacancies is performed at 700°C with a base pressure of 5x10$^{-10}$ mbar. The absence of carbon contamination after annealing is confirmed by in-situ Auger Electron Spectroscopy (AES), see Fig. 3a. Low Energy Electron Diffraction (LEED) performed with the LEEM at 23.26 eV and 41.46 eV, shown in Fig. 3b,c demonstrated a c(2x2) surface reconstruction, already reported in the literature [11] and corresponding to a BaO terminated surface as in the calculations. The oxygen vacancies also help to avoid sample charging. The domain ordering was imaged as a function of electron kinetic energy with respect to the sample surface. The electron reflectivity allows a direct determination of the MEM-LEEM transition energy. It has already been shown that domain surfaces with a positive fixed polarization charge show a MEM-LEEM transition at lower energy than those with a negative fixed polarization charge [12]. The sample was then cycled by moderate annealing through Curie temperature (120°C) into the paraelectric (PE) phase up to 250°C and then cooled back the FE phase at room temperature. The FE-PE-FE cycle was followed by MEM, and the electron reflectivity as a function of kinetic energy with respect to the sample potential, was measured at the beginning and at the end of the FE-PE-FE cycle. The transition energy was extracted by a pixel by pixel fit to the electron reflectivity across the field of view giving a map of the electrostatic potential just above the sample surface.

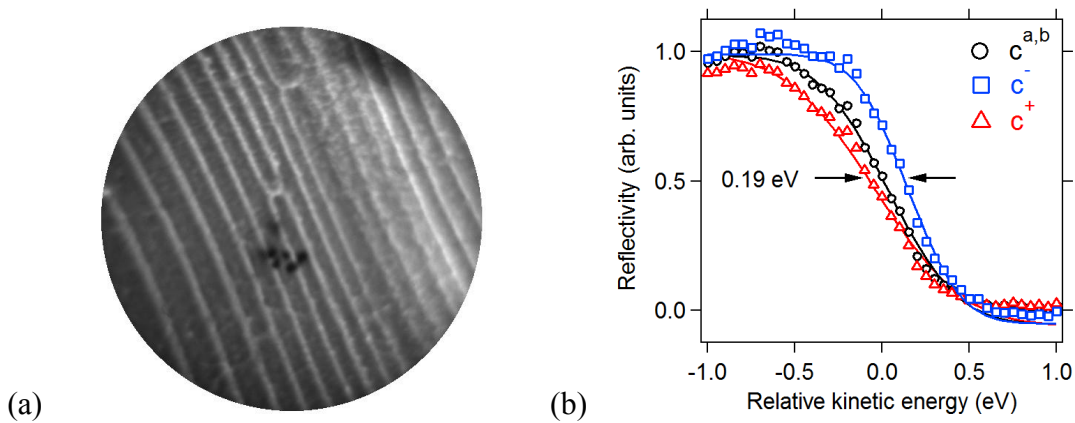

"**Figure 4.** Figure caption. (a) Raw MEM image at a start voltage of -0.15 V showing the maximum contrast in electrostatic potential above the differently polarized domains. The black dots in the centre of the image are defects in the multi-channel plate. (b) Reflectivity curves extracted over the full field of view showing a clear shift in the MEM-LEEM transition for different polarizations."

An example of the raw MEM data is shown in Fig. 4a, the field of view (FoV) is 71 μm. The reflectivity curves extracted from three distinct regions are shown in Fig. 4b and compared with the average reflectivity integrated across the whole FoV. It is clear that different regions on the surface have different MEM-LEEM transitions. The MEM-LEEM transition energy is lowest above $c^+$ domains and shifts 0.19 eV to higher energy above $c^-$ domains. The MEM-LEEM transition maps obtained from the fit to the data are shown in Fig. 5a before and after the FE-PE-FE cycle. Clear domain ordering exists in both FE states. There are three distinct values in histograms of the distribution of MEM-LEEM transitions which can be related to the three possible polarizations: the out of plane $c^+$ and $c^-$ and the in-plane a-type distortions (Fig. 5b). No change in the LEED pattern was observed after the FE-PE-FE cycle and no carbon contamination was observed in AES, therefore domain ordering and oxygen vacancies constitute the two principal contributions to the screening of the depolarizing field.

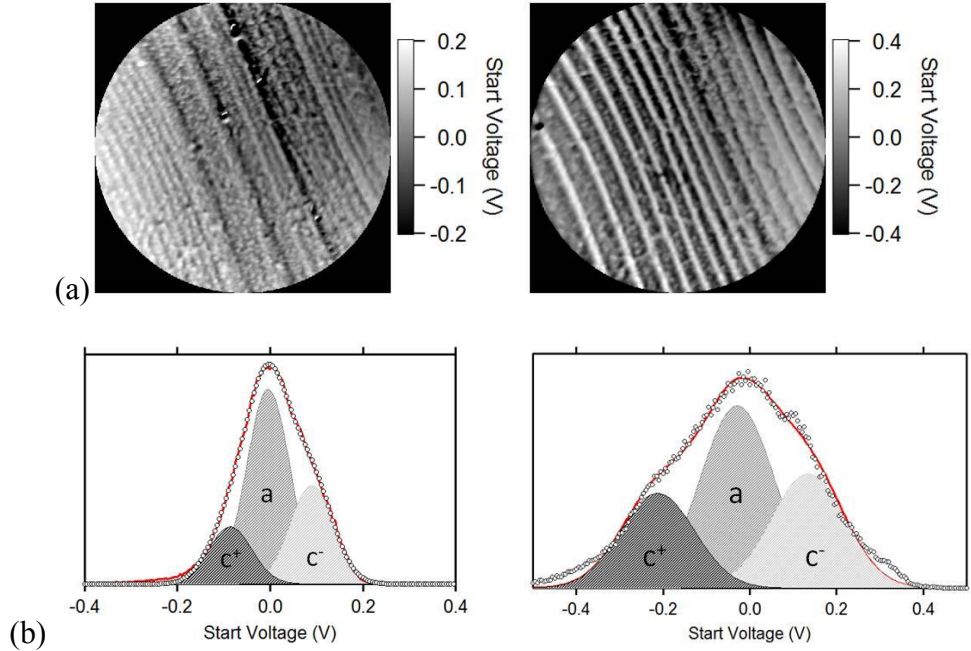

"**Figure 5.** Figure caption. (a) MEM-LEEM transition maps before and after FE-PE-FE cycle showing evolution in domain ordering. (b) Histograms of MEM-LEEM transition values across the field of view showing the existence of three distinct transitions corresponding to $c^+$, $c^-$ and a-type domains."

From Fig. 5b, the potential difference between $c^+/c^-$ MEM-LEEM transitions increases after the FE-PE-FE cycle, as does the proportion of $c^+/c^-$ domains with respect to a-type domains. The possible domain orientations are shown schematically in Fig. 6a. The increase in the difference in electrostatic potential above $c^+/c^-$ domains may be related to the increase in domain width which may reduce screening of the surface charge away from domain walls, whereas the larger proportion of $c^+/c^-$ domains, by favouring 180° domain walls, is consistent with minimizing of the elastic energy of the system. Both before and after the cycle, the $c^-$ domain is more widespread than the $c^+$ domain (Fig. 6b), suggesting that the system energy of the former is lower. The maximum temperature attained during the FE-PE-FE cycle is 250°C, well below the temperatures necessary to create significant oxygen vacancies. Thus, the domain ordering observed cannot be related to changes in the oxygen vacancy concentration, nor to different extrinsic screening since the surface remains clean. Rather, we have evidence that the system domain ordering has evolved to minimize further its total energy, following qualitatively the results of the DFT calculations.

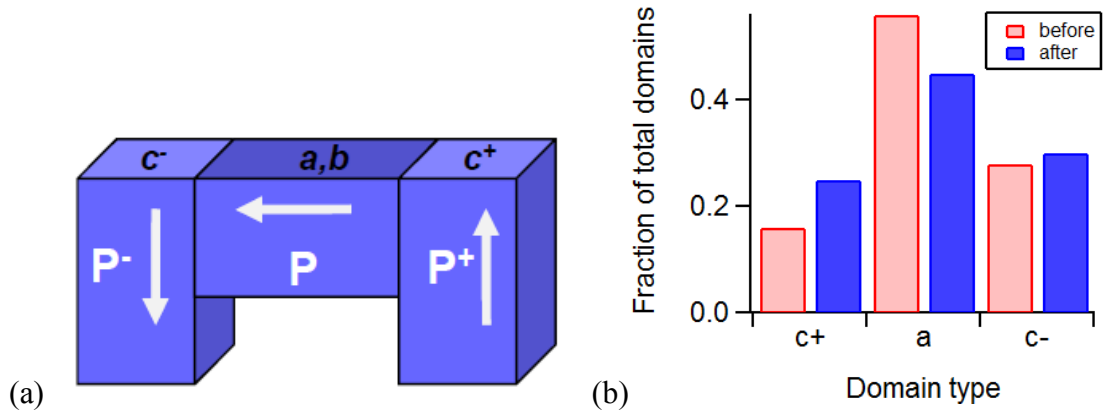

(a) (b)

"**Figure 6.** Figure caption. (a) Schematic view of the three kinds of domains encountered in the present work. (b) Proportion of $c^+$, $c^-$ and a-type domains before and after FE-PE-FE cycle."

**IV - Conclusion**

MEM-LEEM experiments show that $c^-$ domains, i.e. with polarization pointing inwards, are more favorable, at least in BaO-terminated surfaces of barium titanate (001). Our calculations suggest that such dissymmetry between $c^+$ and $c^-$ domains may be due to oxygen vacancies at surfaces, which contribute to stabilize $c^-$ domains.

ACKNOWLEDGMENT. C.M. benefited from post-doctoral funding of the CEA Nanoscience programme. We would like to thank D. Martinotti for technical assistance with the MEM-LEEM experiments at the CEA-Saclay.